\documentclass[twocolumn]{IEEEtran} 
\usepackage{epsf}

\def\BibTeX{{\rm B\kern-.05em{\sc i\kern-.025em b}\kern-.08em
    T\kern-.1667em\lower.7ex\hbox{E}\kern-.125emX}}

\begin{document}

\title{Towards Coherent Neutrino Detection Using 
Low-Background Micropattern Gas Detectors} 

\author{P. Barbeau, J.I. Collar\thanks{ P.B. and J.I.C. are with the
Enrico Fermi Institute and Center for Cosmological 
Physics, University of Chicago, Chicago, IL 60637; J.M. and I.S. are 
with the
Department of Physics, Purdue University, W. Lafayette, IN 47907. E-mail: 
collar@uchicago.edu}, J. Miyamoto, I. Shipsey}

\markboth{IEEE Transactions On Nuclear Science, Vol. XX, No. Y, Month
2003}
{ } 

\maketitle

\begin{abstract}
The detection of low energy neutrinos ($<$ few tens of MeV)
via coherent nuclear scattering remains
a holy grail of
sorts in neutrino physics. This uncontroversial mode 
of interaction is expected to profit from a sizeable increase in cross section 
proportional to neutron number squared in the target 
nucleus, an advantageous feature in view of the small probability of 
interaction via all other channels in this energy region. 
A coherent neutrino detector 
would open the door to many new applications, ranging from the study 
of fundamental neutrino properties to true "neutrino technology". 
Unfortunately, present-day radiation detectors of sufficiently large 
mass ($>$ 1 kg) are not sensitive to sub-keV nuclear recoils like 
those expected from this channel. The advent of Micropattern
Gas Detectors (MPGDs), new technologies originally 
intended for use in High Energy Physics, may soon put an 
end to this impasse. We present first tests of MPGDs fabricated 
with radioclean materials and discuss the approach to assessing their 
sensitivity to these faint signals. Applications are reviewed, 
in particular their use as a safeguard against illegitimate operation 
of nuclear reactors. A first industrial mass production of Gas Electron 
Multipliers (GEMs) is succinctly described.\end{abstract}


\section{Coherent Neutrino Detection: A Technological Challenge}
\PARstart{A} new family of radiation detector designs, generally referred 
to as Micropattern
Gas Detectors
(MPGDs) \cite{reviews} has emerged during the last fifteen years in response to the 
demanding needs (fast counting rate, radiation resistance, high spatial resolution) 
of next-generation High Energy Physics experiments. While 
the specific design varies, 
their common principle is a sizeable voltage drop across 
microstructures immersed in a suitable gas mixture: electrons  
originating from particle ionization in a conversion volume are
multiplied in the microstructures, where  
amplification gains of up 
to $10^{7}$ are obtained. A popular example of a MPGD is the 
MICROMEGAS design (MICROMEsh GAseous Structure), a 
concept recently put forward by Y. Giomataris, G. Charpak 
and collaborators \cite{mms1}. This 
two-stage parallel-plate avalanche chamber consists of a 
100 $\mu$m narrow amplification gap and a large conversion 
region -TPC volumes are possible-, 
separated by a gauze-like electroformed conducting micromesh. Electrons released 
by ionizing particles
in the gas-filled conversion region are drifted towards the 
amplification gap where they multiply in an avalanche process. 
Detectable signals are then induced on anode elements. 
A second example of MPGDs are Gas Electron 
Multipliers (GEMs)\cite{GEMS}, developed at CERN by F. Sauli and 
collaborators: small 
holes (diameter$\sim 80 \mu m$) are photolithographically 
etched on a $\sim 50 \mu m$-thick Kapton film copper-clad on both sides and a 
voltage difference of $\sim 400$ V is generated across the GEM. The high 
density of electric field lines within the perforations induces the 
sought avalanche. An advantage of GEMs is the 
possibility of building multi-stage amplification layers \cite{ian}, allowing 
for very large gains. The high-efficiency detection of single electrons at 
gas pressures of up to 20 atm has been achieved in a variety of MPGDs 
\cite{mms5}. The effective energy threshold in these devices is the 
ionization energy of the gas mixture, i.e., a few tens of eV.
\begin{figure}[tbp]
\epsfxsize = \hsize 
\epsfbox{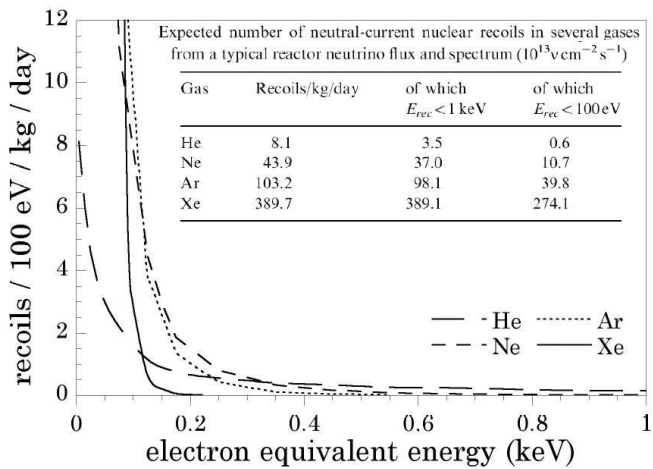}
\caption{Detectable signal in different  
gases from neutral-current nuclear scattering of reactor 
antineutrinos (10$^{13}  \bar{\nu}$ cm$^{-2}$ s$^{-1}$), obtained by 
folding of the differential cross section in [8] with the  
reactor spectrum in [24] and applying quenching factors derived from 
SRIM [25]. 
The tradeoff between endpoint energy and 
rate with increasing atomic mass is evident. 
{\it Table}: total coherent recoil rate in different gases under the same 
conditions.}
\end{figure}
\begin{figure}[tbp]
\epsfxsize = \hsize \epsfbox{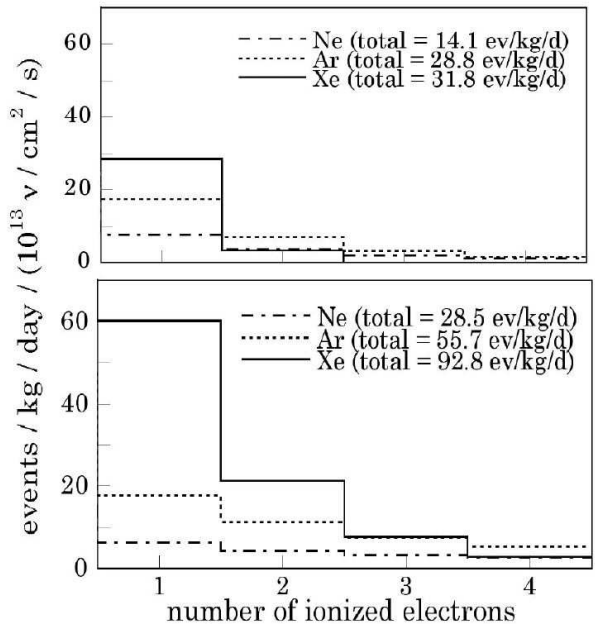}
\caption{Distribution of the number of ionized electrons produced by coherent 
nuclear scattering in 
a gaseous detector exposed to a typical reactor antineutrino flux. This
estimate includes the reactor emission spectrum, differential cross 
section,
a quenching factor derived from Linhard's theory and the mean ionization energy 
of the gas mixtures. {\it Top:} for 
pure noble gases, {\it Bottom:} after addition of a small 
fraction of TMAE vapor which may in principle reduce the ionization 
threshold to $\sim$6 eV. The effect of gas additives such as TMAE or 
TEA on energy threshold is to be
investigated as part of this work.} 
\end{figure}

The  
possibility of exploiting some of the features specific to MPGDs in a 
new 
realm, that of searches for rare-events in neutrino and astroparticle 
physics has been recently examined \cite{meyannis}. 
The properties of these devices (background rejection 
capabilities, demonstrated ability for single-electron detection, 
versatility and simplicity) suggest a means to tackle
a long-standing experimental challenge, the measurement of {\it coherent} neutral-current 
neutrino-nucleus scattering.
An uncontroversial process in the Standard Model, the scattering off 
nuclei of low-energy neutrinos ($< $ few tens of MeV, e.g., reactor 
$\bar{\nu} s$) via the neutral current 
\cite{freedman} remains undetected. The long neutrino wavelength probes 
the entire nucleus, giving rise to a large coherent enhancement in the cross 
section, roughly proportional to neutron number squared \cite{drukier}. 
Using this mode of interaction, it would be possible to speak of 
{\it portable} neutrino detectors: in some experimental conditions 
the expected rates can be as high as several 
hundred recoils/kg/day, by no means a ``rare-event'' situation. 
However, the recoil energy transferred to the target is a few keV at most 
even for 
the lightest nuclei, with only 
a few percent going into ionization (\frenchspacing{Figs. 1,2}).  

The interest in observing this process is not merely academic: a 
neutral-current detector responds the same way to all known neutrino types.
Therefore, the observation of neutrino oscillations in such a device would 
be {\it direct} evidence for a fourth sterile neutrino. These can be invoked if all 
recently observed neutrino anomalies are accepted at face value 
\cite{ellis} and may play an important role as Dark Matter \cite{dolgov}. 
Separately, the cross section for this process is critically dependent on 
neutrino magnetic moment. Agreement with the Standard Model prediction 
would {\it per se} largely improve on the present experimental sensitivity 
to $\mu_{\nu}$\cite{dodd}. In addition to this, a measurement 
of the cross section
would constitute a sensitive probe of the weak nuclear charge,
testing radiative corrections due to new physics above the weak scale 
with a sensitivity comparable to atomic parity violation and accelerator 
experiments. Statistically speaking, 
this can be accomplished in a nuclear reactor
already with a modest 
detector mass and a short exposure
\cite{larry}.
Finally, this coherent mechanism plays a most important role in neutrino 
dynamics in supernovae and neutron stars \cite{freedman}, adding to the 
attraction of a laboratory measurement of this cross section. 
In particular, a measurement 
of the {\it total} (flavor-independent) neutrino flux from a nearby 
supernova using a large enough coherent detector would be of capital importance to 
help clarify the exact oscillation pattern followed by the neutrinos 
in their way to the Earth \cite{beacom}.

\begin{figure}[tbp]
\epsfxsize = \hsize \epsfbox{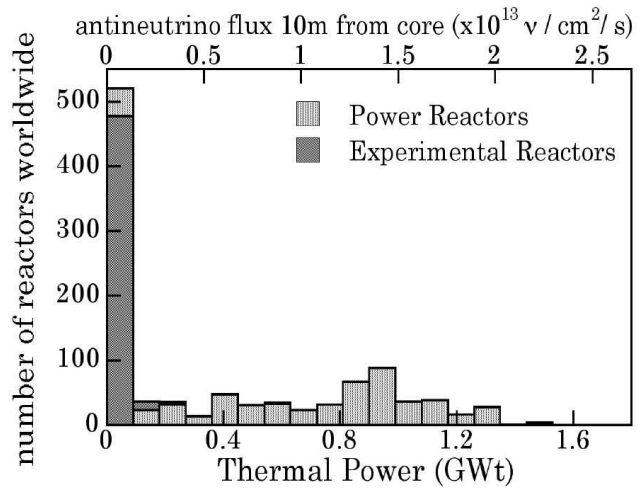}
\caption{Power and approximate 
antineutrino flux distribution of worldwide nuclear 
reactors, extracted from the databases in [26]. 
Proposals to monitor illicit reactor activity using conventional neutrino 
detectors (large liquid scintillator tanks) seem
insufficient for this purpose:
their sensitivity under realistic conditions would be adequate only for 
reactor powers larger than $\sim 3$ GWt and require the construction 
of underground infrastructure ($\sim$ 6 m.w.e.) close to reactor 
cores [16]. 
Detectors based on coherent scattering may be able to improve this 
situation in the near future.} 
\end{figure}
\begin{figure}[tbp]
\epsfxsize = \hsize \epsfbox{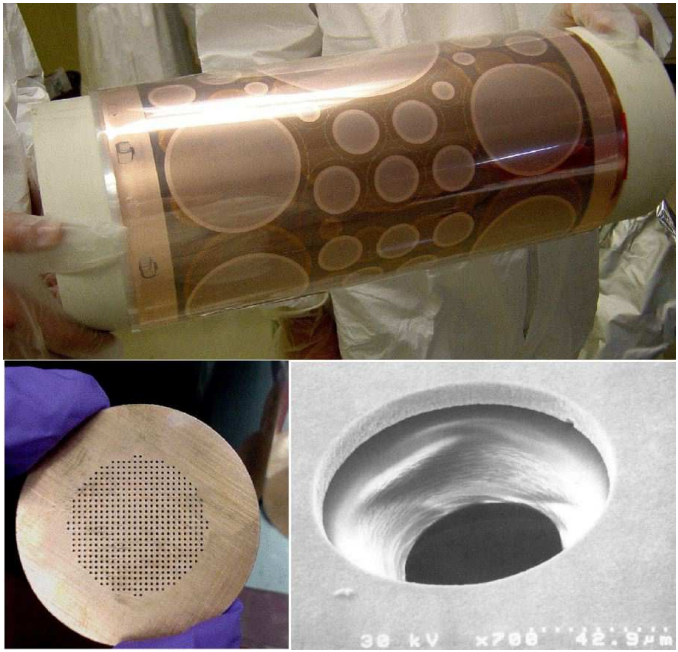}
\caption{{\it Top}: A first mass-production of GEMs using 
3M's Microflex adhesiveless reel-to-reel process. 
The roll in the figure contains 35 panels 
of 33 GEM elements each. Any GEM pattern up to 12''x12'' 
can be produced.
Perforations to facilitate detachment are visible around each element.
{\it Bottom right}: SEM photograph of one of these GEMs 
(hole diameter 80 $\mu$m, 
pitch 140 $\mu$m). {\it Bottom left}: LEMs produced at 
EFI using automated micromachining on low background laminates 
(OFHC copper plated directly onto virgin Teflon).} 
\end{figure}
\begin{figure}[tbp]
\epsfxsize = \hsize \epsfbox{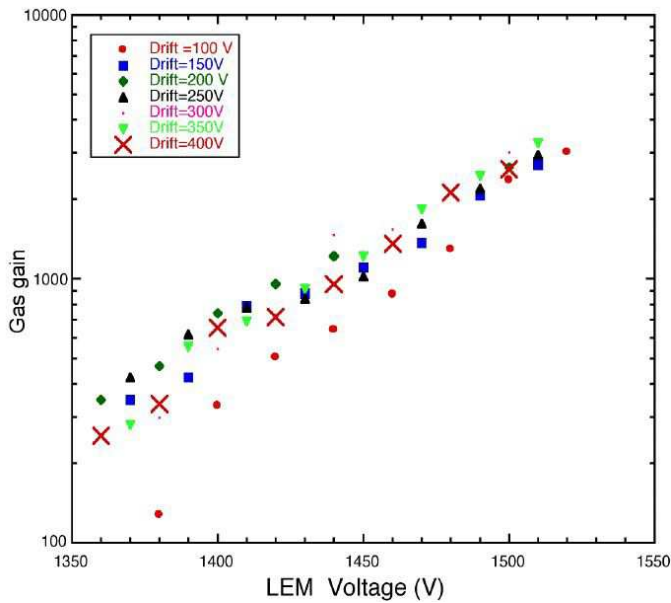}
\caption{Preliminary tests of gas gain in 
a single LEM  (Ar:DME(9:1) at 1 atm) using a $^{55}$Fe uncollimated 
source. The drift cathode was grounded and the LEM held at a 
positive potential (drift distance = 0.5 cm). 
A multi-layer LEM structure should provide enough amplification to detect 
single electrons at moderate gas overpressures. We plan to investigate 
the dependence of the proportional-scintillation light yield on operating 
pressure as a possible mechanism to increase gas density while 
maintaining single-electron sensitivity.} 
\end{figure}

Until now, no existing device 
had met the mass and energy threshold requirements involved in this 
measurement, even though unrealized cryogenic proposals abound \cite{blas}. 
A considerable fraction of the neutrino signal in a 
reactor experiment is 
nevertheless expected above MPGD energy thresholds (Fig. 2).
Structurally simple 
MPGD-based coherent neutrino detectors would open the door 
to more mundane but no less 
important applications than those listed above (``neutrino 
technology''? \cite{leocastle}): 
for instance, nonintrusive monitoring of nuclear reactors against illegitimate uses 
(e.g., fuel rod diversion, unauthorized production of weapon-grade material)
with a compact device, potentially improving on existing proposals 
that rely on standard neutrino detectors and processes 
\cite{bernstein} (Fig. 3). 

\section{Present Status and Immediate Plans}

We have recently commenced fabrication 
and characterization of radioclean MPGDs with 
a first goal of 
coherent neutrino detection while keeping in mind other 
possible applications of the 
same devices, e.g., Weakly Interacting Massive Particle (WIMP) searches.
Three techniques are currently being pursued in parallel: use of a
Micromegas backpanel as a proportional-scintillation reflector,
multi-stage Gas Electron Multipliers (GEMs) and  
Large Electron Multipliers (LEMs). The last are 
similar to GEMs but with 
all dimensions increased by a factor of ten.
The use of LEMs may be advantageous in applications like the present 
one where no spatial 
information is required and only modest energy resolution is needed. 
As a trade-off they offer a larger  
resistance against discharge-induced damage than GEMs, due to their reduced 
capacitance, and the simplicity that comes with their being self-supporting
(no careful mounting and stretching on a frame is needed as in the case 
of GEMs). 
LEMs have been previously 
considered in the context of large TPCs and WIMP detectors \cite{Jeanneret}.
Several LEM prototypes ranging in thickness from 0.25 to 0.75 mm (Fig. 4)
have been
micromachined at
the Enrico Fermi Institute (EFI) from low-activity materials (10 
$\mu$m 
OFHC copper plated directly onto virgin Teflon) 
and have undergone satisfactory preliminary tests 
at Purdue. First measurements on  
single LEMs are encouraging:
only a modest increase in voltage is needed to produce gains similar 
to GEMs (\frenchspacing{Fig. 5}), most probably due to the longer avalanche 
regions. They 
nevertheless exhibit a diminished 
energy resolution in comparison to GEMs (\frenchspacing{Fig. 
6}). While more detailed studies are underway, we can hypothesize 
that this effect is due to a large fraction of primary ionizations 
taking place within the LEM holes (in these calibrations 
the conversion volume was small, a 0.5 cm drift distance). 
If this is the case, the resolution is
expected to improve for larger TPC volumes. As expected, the rise 
time of the signal is also slower than in GEMs, an effect 
unimportant for most low-counting rate applications (\frenchspacing{Fig. 
7}). It is also observed
that the leakage current across these LEMs is of only a few pA @ 
2000 V, whereas typical GEMs exhibit values in the few nA @ 
$\sim$500 V (the volume resistivity of Teflon is 
$\sim\!10^{18} $ ohm$\cdot$cm while this is 
$\sim\!2.3\times 10^{16} $ ohm$\cdot$cm
in Kapton).
A second production of LEMs using a polyetheretherketone substrate is 
underway (polyetheretherketone exhibits an even lower outgassing 
than Teflon).
\begin{figure}[tbp]
\epsfxsize = \hsize \epsfbox{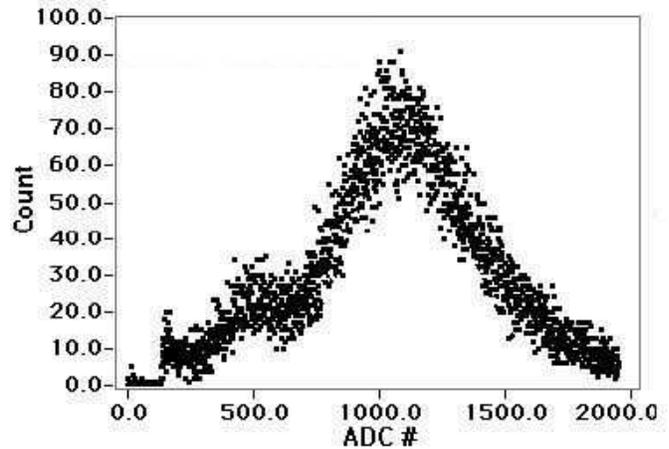}
\caption{Observed resolution in a single LEM under horizontal 
irradiation using an uncollimated $^{55}$Fe source (10 cm$^{2}$ active area, 
0.5 cm drift distance, 1 atm Ar:DME(9:1), gain = 1000). The Ar escape 
peak is visible.} 
\end{figure}
\begin{figure}[tbp]
\epsfxsize = \hsize \epsfbox{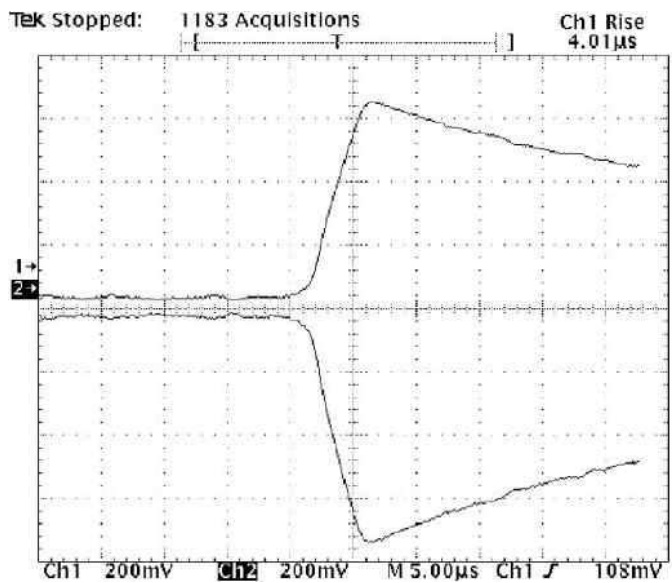}
\caption{Signal development in a single LEM under $^{55}$Fe 
irradiation. The top (bottom) trace corresponds to the anode (cathode) 
signal. Timing and amplitude are identical. A slow $\sim$ 4 $\mu$s rise 
time is observed (some 60 times slower than in a typical GEM), as 
expected from the large (800 $\mu$m) avalanche regions.} 
\end{figure}
\begin{figure}[tbp]
\epsfxsize = \hsize \epsfbox{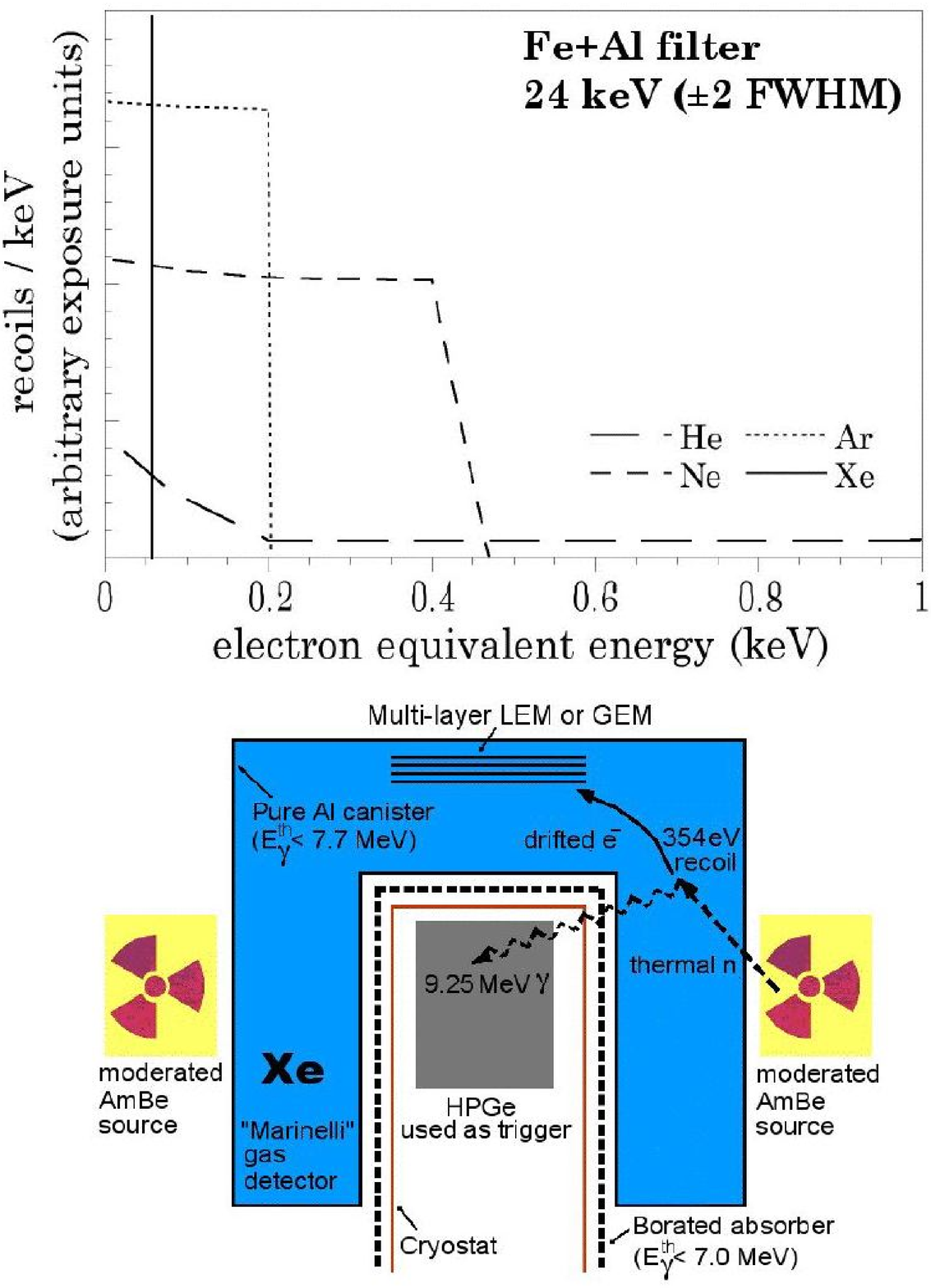}
\caption{{\it Top:} Recoil signals (energy lost to ionization) 
expected in different target gases
from a filtered (Fe+Al) neutron beam of 24 keV (2 keV 
FWHM) using the IPNS facility at Argonne National Laboratory. 
The energy distribution mimics that expected from reactor 
antineutrinos (Fig.1). Other neutron energies to be used in these 
calibrations are 55 keV (Si+S) and 144 
keV (Si+Ti) [20]. The distribution of recoils in the figure is obtained from 
SPECTER [27]. {\it Bottom:} A table-top setup able to produce 
low-energy monochromatic recoils 
in the range 140-350 eV in 
Xe (see text).} 
\end{figure}

Separately, sixty 16''x16'' GEM panels 
(for a total of 1980 GEM elements) 
have been produced in collaboration with 3M \cite{3M} using their proprietary 
reel-to-reel FLEX technology \cite{3M2} (\frenchspacing{Fig. 4}). 
This is the first instance of GEM 
industrial mass production: until now GEMs have been 
available exclusively from CERN, generally in 
small surface areas most suitable for R\&D. Two different techniques 
(additive and subtractive copper cladding) have 
been tested, with a third one under production. A large variety of 
finishings and treatments is possible from 3M's production line: for 
instance, the periphery of each GEM element within a panel 
can be perforated for easy detachment. Any GEM pattern is possible, up 
to a 12''x12'' size. At the time of this writing 
the first batch is undergoing testing at EFI. Their characterization 
will be the subject of an upcoming publication. 

The short-term physics objectives are:

$\bullet$ A calibration facility to provide  monochromatic (filtered)
neutron \cite{nbeams} beams able to produce recoils almost identical to those 
expected from reactor antineutrinos (\frenchspacing{Fig. 
8}) is to be built at the 
Intense Pulsed Neutron Source (Argonne National 
Laboratory).
These measurements will provide not only a convincing proof of the 
ability of MPGDs to detect these low-energy signals, but also a chance 
to characterize the low-energy quenching factors and 
thresholds for different gas 
mixtures
as well as the attainable gain as a function of gas pressure. 
This information is of the utmost importance for the interpretation of a 
subsequent 
neutrino experiment. The same facility can be later employed to 
characterize WIMP detectors. A second calibration setup 
presently under construction uses well-defined 
monochromatic daughter recoils from the 
Xe(n$_{thermal},\gamma)$ 
reaction, ranging in energy from 140 eV to 350 eV  (\frenchspacing{Fig. 
8}). Monte Carlo simulations show that
a careful selection of materials can ensure a high 
signal-to-noise ratio. Thermal neutron absorption has been used before to
study quenching factors in Ge for recoil energies down to 250 eV \cite{jones}.

$\bullet$ An interesting intermediate physics result is expected 
from measurements of intrinsic detector backgrounds, to take place at 
a depth of 
60 m.w.e. in the low-background laboratory at EFI. A four-liter OFHC 
Cu prototype is under construction for this purpose. This unique
combination of shielding against cosmic rays, sizeable 
target mass ($\sim 80$ g) 
and ultra-low energy threshold 
should return an improvement of several orders of magnitude 
on the present experimental sensitivity to a slow 
solar-bound WIMP population \cite{solar} and to 
recently proposed non-pointlike dark 
matter particle candidates \cite{graciela}. While the 
nature of radioactive backgrounds  
below $\sim$1 keV is a true {\it terra incognita}
for large devices, experience in WIMP 
detector development indicates that no sudden rise is 
expected in this energy region 
from known natural sources. Low-energy neutron recoils and recoiling 
daughters 
from (n$_{thermal},\gamma)$ can be controlled with layers of 
moderating and 
absorbing shielding. Degraded $\alpha$ and $\beta$ radiations from 
surfaces can be kept to a minimum using radioclean materials in 
the detector construction. Similarly, if the need ever arises, 
it should be possible to reduce 
spurious single-electron emission from 
Malter and field effects down to a negligible level via surface 
treatment (as in accelerating RF-gun cavities) 
and rigorous control of gas composition and purity.
\vskip0.2cm

The achievement of these short-term goals will take this project to the 
point where a first measurement of this exciting mode of neutrino 
interaction can 
be performed in a nuclear reactor.

\nocite{*}
\bibliographystyle{IEEE}

%

%

\end{document}